\documentclass[aps,prb,twocolumn,groupedaddress,showpacs,showkeys,amsmath,amssymb]{revtex4}

\usepackage{amsfonts}
\usepackage{amssymb,amsmath}
\usepackage{graphicx}
\usepackage{dcolumn}

\begin{document}

\title{Magnetic-field induced spin-Peierls instability
       in strongly frustrated quantum spin lattices}

\author{Johannes Richter$^\dagger$,
        Oleg Derzhko$^\dagger$\footnote
        {On leave of absence from
        the Institute for Condensed Matter Physics,
        National Academy of Sciences of Ukraine,
        1 Svientsitskii Street, L'viv-11, 79011, Ukraine}
        and
        J\"{o}rg Schulenburg$^\ddagger$}

\affiliation{$^\dagger$Institut f\"{u}r Theoretische Physik,
             Universit\"{a}t Magdeburg,
             P.O. Box 4120, D-39016 Magdeburg, Germany\\
             $^\ddagger$Universit\"{a}tsrechenzentrum,
             Universit\"{a}t Magdeburg,
             P.O. Box 4120, D-39016 Magdeburg, Germany}

\date{\today}



\begin{abstract}
For a class of frustrated antiferromagnetic spin lattices
(in particular,
the square-kagom\'{e} and kagom\'{e} lattices)
we discuss the impact of recently discovered exact eigenstates
on the stability of the lattice against distortions.
These eigenstates consist of independent localized magnons
embedded in a ferromagnetic environment
and become ground states in high magnetic fields.
For appropriate lattice distortions
fitting to the structure of the localized magnons
the  lowering of magnetic energy can be calculated exactly
and is proportional to the displacement of atoms
leading to a spin-Peierls lattice instability.
Since these localized states are present only for high magnetic fields,
this instability might be driven by magnetic field.
The hysteresis of the spin-Peierls transition is also discussed.
\end{abstract}

\maketitle

Antiferromagnetically interacting spin-$\frac{1}{2}$ systems
on geometrically frustrated lattices
have attracted much attention during last years.
Such systems  have rich phase diagrams
exhibiting a number of unusual quantum phases \cite{01,02}.
A striking example is the kagom\'{e} lattice antiferromagnet
having a liquid-like ground state
with a gap for magnetic excitations
and a huge number of singlet states below the first triplet state
(see e.g.
Ref. \onlinecite{waldtmann98} and references therein).
Another intriguing example
is the structural phase transition in spin systems
driven by magnetoelastic coupling (spin-Peierls instability)
observed, e.g., in CuGeO$_3$ \cite{CuGeO}.
Frustrating interactions may also provide a route
to generating fractional phases in two dimensions
which manifest itself in the dynamic correlations
probed by inelastic neutron scattering experiments \cite{03}.

In the presence of an external magnetic field
frustrated quantum spin systems exhibit
a number of unusual properties.
In particular,
plateaus and jumps can be observed
in the zero-temperature magnetization curve
for such models \cite{01,04}.
The theoretical investigation of exotic magnetization curves
has been additionally stimulated
by the experimental observation of plateaus
e.g.
in CsCuCl$_3$ \cite{nojiri88}
or
SrCu$_2$(BO$_3$)$_2$ \cite{kageyama99}.

Though the treatment of quantum spin systems
often becomes more complicated
if frustration is present,
in some exceptional cases frustration is crucial
to find simple ground states
of product form \cite{majumdar_ghosh,ivanov_richter}.
Recently,
for a wide class of frustrated spin lattices
exact eigenstates consisting of independent localized magnons
in a ferromagnetic environment have been found \cite{05}.
They may become ground states if a strong magnetic field is applied
and lead to a macroscopic jump
in the zero-temperature magnetization curve just below saturation.

In the present paper
we examine the stability of antiferromagnetic spin lattices
hosting independent localized magnons
with respect to  lattice distortions 
through a magnetoelastic mechanism.
We are able to present rigorous  analytical results
completed by large-scale exact diagonalization data
for lattices up to $N=54$ sites.
We discuss
the field-tuned changes of the ground-state properties
owing to a coupling between spin and lattice degrees of freedom.

The effect of a magnetoelastic coupling in frustrated antiferromagnets
is currently widely discussed.
First of all
the investigation of the Peierls phenomena
in frustrated 1D spin systems is in the focus
(see, for instance, Ref. \onlinecite{07} and references therein).
But also in 2D and 3D quantum spin systems
lattice instabilities breaking the translational symmetry
are reported.
The frustration-driven structural distortions
in the spin-$\frac{1}{2}$ square-lattice $J_1$-$J_2$ Heisenberg antiferromagnet
studied in Ref. \onlinecite{08}
might be relevant for Li$_2$VOSiO$_4$ and VOMoO$_4$ \cite{08,09}.
For 3D frustrated antiferromagnets
containing corner-sharing tetrahedra
one discusses several examples for lattice distortions.
Inelastic magnetic neutron scattering on ZnCr$_2$O$_4$
revealed that a lattice distortion can lower the energy
driving the spin system into an ordered phase \cite{10}.
NMR investigation
of the three-dimensional pyrochlore antiferromagnet
Y$_2$Mo$_2$O$_7$
gives evidence
for discrete lattice distortions
which reduce the energy \cite{11}.
A lifting of a macroscopic ground-state degeneracy
of frustrated magnets
through a coupling between spin and lattice degrees of freedom
in pyrochlore antiferromagnets
was studied in Refs. \onlinecite{12,13},
which may have relevance
to some antiferromagnetic compounds with pyrochlore structure \cite{13}.

In all those studies
the lattice instability is discussed  at zero field.
As pointed out already in the late seventies \cite{gross}
a magnetic field may act against the spin-Peierls transition
and might favor a uniform or incommensurate phase.
In contrast to those findings,
in the present  paper we discuss magnetic systems
for which the magnetic field 
is essential for the occurrence of the lattice instability. 

\begin{figure}[t]
   \begin{flushleft}
       \leavevmode
     \vspace{1.25cm}
       \epsfxsize=4.0cm
       \epsffile{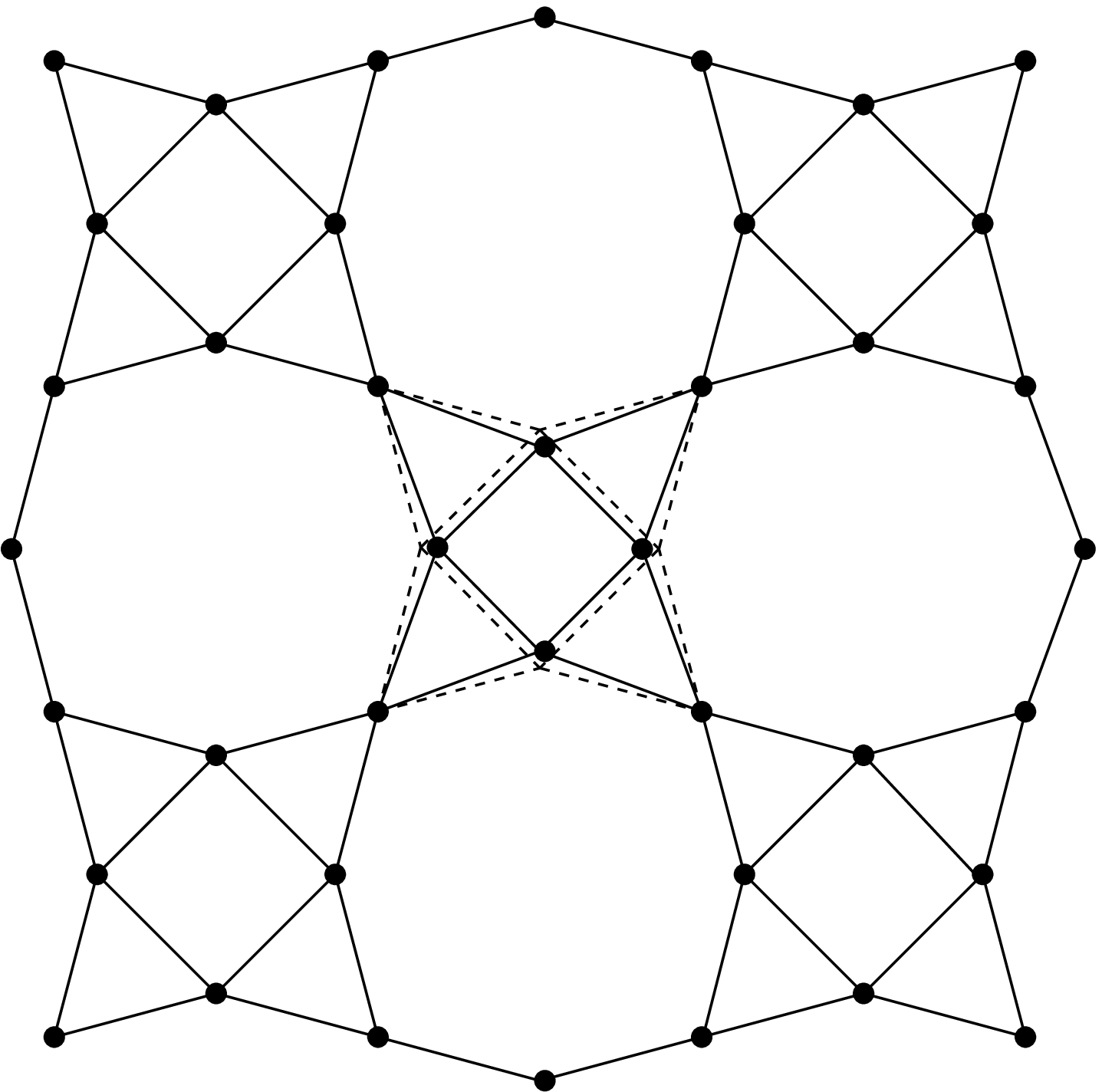}
   \end{flushleft}
     \vspace{-5.75cm}
   \begin{flushright}
       \leavevmode
       \epsfxsize=4.0cm
       \epsffile{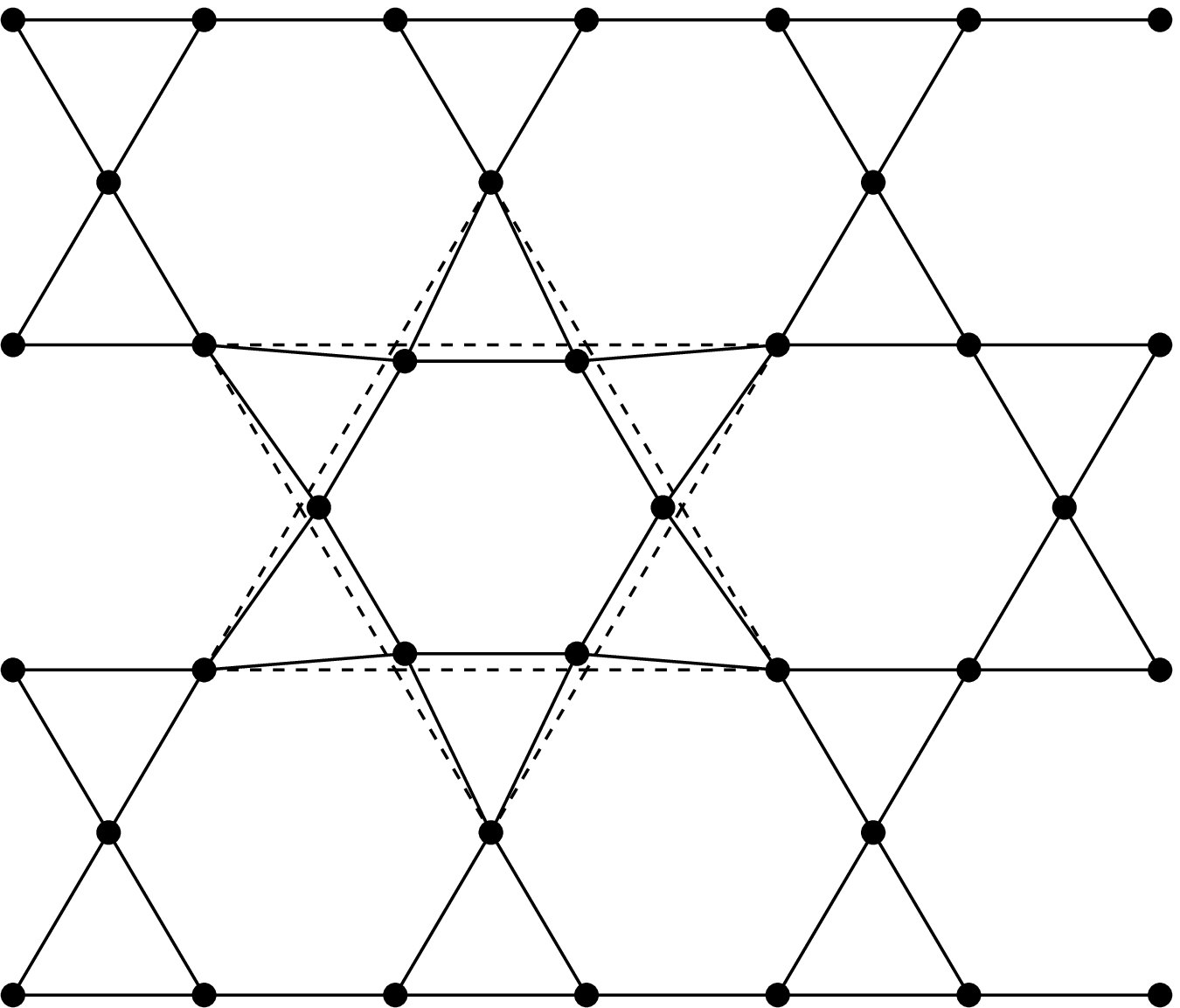}
   \end{flushright}
\caption
{Square-kagom\'{e} lattice with one distorted square
(left)
and
kagom\'{e} lattice with one distorted hexagon
(right)
which can host localized magnons.
The parts of the lattices before distortions
are shown by dashed lines.
All bonds in the lattices before distortions have the same length.
\label{fig01}}
\end{figure}

To be specific,
we consider two geometrically frustrated lattices,
namely,
the square-kagom\'{e} lattice
(Fig. \ref{fig01}, left)
\input epsf
and
the kagom\'{e} lattice
(Fig. \ref{fig01}, right).
The ground state and low-temperature thermodynamics
for the Heisenberg antiferromagnet on both lattices
are subjects of intensive discussions
(see e.g. Refs. \onlinecite{01,02,waldtmann98,14,tomczak}).
The Hamiltonian
of $N$ quantum ($s=\frac{1}{2}$) spins
reads
\begin{eqnarray}
\label{01}
H=
\sum_{(nm)}J_{nm}\left(\frac{1}{2}\left(s_n^+s_m^-+s_n^-s_m^+\right)
+\Delta s_n^zs_m^z\right)
-hS^z.
\end{eqnarray}
Here
the sum runs over the bonds (edges)
which connect the sites (vertices)
occupied by  spins
for the lattice under consideration,
$J_{nm}>0$ are the antiferromagnetic exchange constants
between the sites $n$ and $m$,
$\Delta\ge 0$
is the anisotropy parameter
(in most cases throughout this paper $\Delta=1$),
$h$ is the external magnetic field
and
$S^z=\sum_ns_n^z$ is the $z$-component of the total spin.
We assume that all bonds in the lattice
without distortion
have the same length
and hence all exchange constants have the same value $J$.

From Ref. \onlinecite{05} we know
that independent localized one-magnon states
embedded in a ferromagnetic background
are exact eigenstates of the Hamiltonian (\ref{01})
for the considered models.
More specifically,
by direct computation one can check\cite{05}
that
\begin{eqnarray}
\label{02}
|1\rangle&=&\frac{1}{2}\sum_{\stackrel{i=1}{i \in square}}^4 (-1)^i 
s_i^-\vert FM\rangle, 
\\
\label{03}
|1\rangle&=&\frac{1}{\sqrt{6}}\sum_{\stackrel{i=1}{i \in hexagon}}^6 (-1)^i
s_i^-\vert FM\rangle 
\end{eqnarray}
are the one-magnon eigenstates of the Hamiltonian (\ref{01})
on the square-kagom\'{e} and kagom\'{e} lattices, where 
$\vert FM\rangle$
stands for the embedding fully polarized ferromagnetic environment.
The corresponding energies 
($h=0$) of the one-magnon states (\ref{02}) and (\ref{03}) are
\begin{eqnarray}
\label{04}
-J+J+\left(2N-12\right)\frac{1}{4}J,\\
\label{05}
-\frac{1}{2}J+2J+\left(2N-18\right)\frac{1}{4}J .
\end{eqnarray}
The magnons (\ref{02}) or (\ref{03})
are trapped (localized) on a square or on a hexagon,
respectively.
We separate explicitly in (\ref{04}), (\ref{05})
the contributions to the energy
from those bonds which form a magnon trapping cell (first terms),
from the bonds connecting this cell with the environment (second terms)
and
from the ferromagnetic environment (third terms).
One can proceed to fill the lattices with $n > 1$ localized magnons;
the state with maximum filling (``magnon crystal'') has
$n=n_{\max}$ independent localized magnons
with
$n_{\max}=\frac{1}{6}N$
and
$\frac{1}{9}N$
for the square-kagom\'{e} and kagom\'{e} lattices,
respectively \cite{05}.
Since $S^z$ commutes with the Hamiltonian (\ref{01}),
the energy in the presence of an external field $h\ne 0$,
$E(S^z,h)$,
can be obtained from the energy without field,
$E(S^z)$,
through the relation
$E(S^z,h)=E(S^z)-hS^z.$
Since each magnon carries one down spin,
a localized magnon state with $n$ independent magnons
has $S^z=\frac{1}{2}N-n$ and consequently 
Eqs. (\ref{04}), (\ref{05}) give $E(S^z=\frac{1}{2}N-1)$
for the corresponding systems.

Under quite general assumptions it was proved \cite{15}
that these localized magnon states have lowest energies
in the corresponding sector of total $S^z$.
As a result,
these states
become ground states
in a strong magnetic field.
More specifically,
the ground-state energy in the presence of a field
is given by
$E_0(h)=E_{\min}(S^z)-hS^z$
and the ground-state magnetization $S^z$
is determined from the equation
$h=E_{\min}(S^z)-E_{\min}(S^z-1)$.
Since for $S^z=\frac{1}{2}N,\ldots,\frac{1}{2}N-n_{\max}$ 
(i.e., $0\le n \le n_{\max}$)
the localized magnon states
are the lowest states, one has
\begin{eqnarray}
\label{06a}
E_{\min}(S^z)
=\frac{1}{2}NJ-3nJ
=-NJ+3JS^z
\end{eqnarray}
for both models.
Due to the linear relation between $E_{\min}$ and $S^z$
one has a complete degeneracy
of all localized magnon states
at $h=h_1$,
i.e. the energy is $-NJ$
at $h=h_1$
for all $\frac{1}{2}N-n_{\max}\le S^z\le \frac{1}{2}N$,
where
$h_1=3J$ is the saturation field
(identical for both models).
Consequently,
the zero-temperature magnetization $S^z$ jumps
between the saturation value $\frac{1}{2}N$
and the value
$\frac{1}{3}N$ ($\frac{7}{18}N$) for the square-kagom\'{e} (kagom\'{e})
lattice.
The effects of the localized magnon states
become irrelevant if the spins become classical
($s\to\infty$).

We want to check the lattice stability of the considered systems
with respect to a spin-Peierls mechanism.
For this purpose
we assume
a small lattice deformation which preserves the symmetry of the cell
which hosts the localized magnon
(in this case
the independent localized magnon states
remain the exact eigenstates)
and analyze the change in the total energy
(which consists of the magnetic and elastic parts)
to reveal whether the deformation is favorable or not.
To find a  favorable deformation
one needs optimal gain in magnetic energy.
For that we use the circumstance
that due to the localized nature of the magnons
we have an inhomogeneous distribution
of nearest-neighbor (NN) spin-spin correlations
$\langle { \bf s}_i {\bf s}_j\rangle$ \cite{05}.
In case that 
one magnon is distributed uniformly over the lattice the deviation of the 
NN correlation from the ferromagnetic value, 
$\langle{\bf{s}}_i{\bf{s}}_j\rangle-\frac{1}{4}$,
is of the order
$\frac{1}{N}$.
On the other hand
for a localized magnon (\ref{03})
[(\ref{02})]
we have along the hexagon [square] hosting the localized magnon
actually a negative NN correlation
$\langle{\bf{s}}_i{\bf{s}}_j\rangle
=-\frac{1}{12}\;
[-\frac{1}{4}]$ and all
other correlations are positive.
Hence
a deformation with optimal gain in magnetic energy
shall lead
to an increase of antiferromagnetic bonds
on the square (hexagon) and to a decrease of the bonds on the attaching
triangles.
The corresponding deformations
are shown in Fig. \ref{fig01}.
For the square-kagom\'{e} lattice (Fig. \ref{fig01}, left) the
deformations  
lead to the following changes in the exchange interactions:
$J\to \left(1+\sqrt{2}\delta\right)J$
(along the edges of the square) and 
$J\to \left(1-\frac{1}{4}\sqrt{2}\left(\sqrt{3}-1\right)\delta\right)J$
(along the two edges of the triangles attached to the square),
where the quantity
$\delta$
is proportional to the displacement of the atoms and the change in the
exchange integrals due lattice distortions is taken into account
in first order in $\delta$.
For the kagom\'{e} lattice (Fig. \ref{fig01}, right)
one has
$J\to \left(1+\delta\right)J$
(along the edges of the hexagon)
and
$J\to \left(1-\frac{1}{2}\delta\right)J$
(along the two edges of the triangles attached to the hexagon).
The magnetic energies (\ref{04}) and (\ref{05})
are lowered by distortions
and become
$\frac{1}{2}NJ-3J
-\frac{1}{4}\sqrt{2}\left(3+\sqrt{3}\right)\delta J$
and
$\frac{1}{2}NJ-3J
-\frac{3}{2}\delta J$,
respectively.
This is in competition with the increase of the elastic energy,
which is given  in harmonic approximation by
$2\left(6-\sqrt{3}\right)\alpha\delta^2$ (square-kagom\'{e})
and
by $9\alpha\delta^2$ (kagom\'{e}).
The parameter
$\alpha$
is proportional to the elastic constant of the lattice.
The change of total energies due to distortions read
\begin{eqnarray}
\label{07}
-\frac{1}{4}\sqrt{2}\left(3+\sqrt{3}\right)\delta J
+2\left(6-\sqrt{3}\right)\alpha\delta^2,
\\
\label{08}
-\frac{3}{2} \delta J+9\alpha\delta^2,
\end{eqnarray}
where for $n$ independent localized magnons
with $n$ related distortions
these results have to be multiplied by $n$.
Minimal total energy is obtained for
$\delta=\delta^\star
=\frac{\sqrt{2}\left(3+\sqrt{3}\right)}{16\left(6-\sqrt{3}\right)}
\frac{J}{\alpha}$
for the square-kagom\'{e}
and
$\delta=\delta^\star
=\frac{1}{12}\frac{J}{\alpha}$
for the kagom\'{e} lattice.

We have considered only a special class of lattice deformations
(under which
the independent localized one-magnon states survive).
Although we are not able to prove rigorously
that these lattice deformations are the most favorable,
we have presented above nonrigorous arguments
that these deformations
take advantage of the localized magnons in an optimal way.
However,
we have rigorously shown
that
{\it{there exist lattice deformations
which yield a gain in the total energy
for large values of $S^z$}}
leading to a spin-Peierls  instability of the lattice
for an appropriate (large) magnetic field $h_1$.

To discuss the scenario of spin-Peierls instability more specifically
we consider a magnetic field above the saturation field $h_1$.
For the corresponding
fully polarized ferromagnetic state a lattice distortion is not favorable.
Decreasing $h$ till $h_1$
the homogeneous ferromagnetic state
transforms into the ``distorted magnon crystal'';
this transformation
is accompanied by the aforementioned magnetization jump.
On the basis of general arguments \cite{17,18}
we expect
that the ``magnon crystal'' state has gapped excitations
and the system exhibits a magnetization plateau
between $h_1$ and $h_2 < h_1$
at $S^z=\frac{1}{2}N-n_{\max}$.
To support this statement
we calculate the plateau width $\Delta h = h_1-h_2$
for finite systems
of $N=27,36,45,54$
(kagom\'{e})
and
$N=24,30,48,54$
(square-kagom\'{e})
for the undistorted lattice,
where $h_2$ is obtained by
$h_2
=E_{\min}(S^z=\frac{1}{2}N-n_{\max})
-E_{\min}(S^z=\frac{1}{2}N-n_{\max}-1)$.
Using a $\frac{1}{N}$ finite-size extrapolation
we find indeed evidence
for a finite $\Delta h =0.07 J$ ($\Delta h = 0.33 J$)
for the kagom\'{e} (square-kagom\'{e}) lattice
in the thermodynamic limit.
This plateau width might be enlarged by distortions (see below).

Now the question arises
whether the lattice distortion under consideration
is stable below this plateau,
i.e., for
$S^z < \frac{1}{2}N-n_{\max}$.
We are not able to give a rigorous answer
but can discuss the question again
for finite systems
of size
$N=24,30,48,54$  (square-kagom\'{e})
and
$N=18, 27, 36, 45, 54$ (kagom\'{e})
with $n_{\max}$ distorted squares/hexagons.
We calculate the magnetic energy
for zero and small distortion parameter $\delta$
for different values of $S^z$.
Adopting for the magnetic energy
the ansatz
\begin{eqnarray}
\label{09}
E_{\min}(S^z,\delta)=E_{\min}(S^z,0)+ A\delta^p
\end{eqnarray}
and taking $\delta$ of the order of $10^{-4}$
we can estimate the exponent $p$ from the numerical results.
Evidently,
the lattice may become unstable if $p<2$
(and, of course, $A<0$)
whereas
$p\ge 2$ indicates lattice stability.

Of course,
the numerical results reproduce the analytical findings
reported above for  $S^z = \frac{1}{2}N-n_{\max}$.
More interesting is the sector of $S^z$ just below,
i.e.
$S^z=\frac{1}{2}N-n_{\max}-1$.
Remarkably
both lattices behave differently
as $h$ becomes smaller than $h_2$.
For the finite square-kagom\'{e} lattices considered
we find
$p=1.001,\;1.000,\;1.000,\;1.000$
for
$N=24,30,48,54$,
respectively,
if $S^z=\frac{1}{2}N-n_{\max}-1$.
Moreover,
$p$ remains equal to 1 for smaller $S^z$.
Therefore,
we conclude
that the distorted square-kagom\'{e} lattice
remains stable for $S^z<\frac{1}{2}N-n_{\max}$.
On the other hand,
for finite  kagom\'{e} lattices
we obtain
$p=2.000,\;1.002,\;2.000,\;1.998,\;2.055$
for
$N=18,27,36,45,54$,
respectively,
if $S^z=\frac{1}{2}N-n_{\max}-1$.
The ``outrider'' for $N=27$
may be attributed to finite-size effects,
indeed
we have $p=1.994$ for the next lower $S^z=\frac{1}{2}N-n_{\max}-2$.
Moreover,
$p$ remains about 2 for smaller $S^z$.
We interpret small deviations from 2 also as finite-size effects
and conclude
that the spin-Peierls instability in the kagom\'{e} lattice
(within the adopted ansatz for the lattice deformation)
is favorable only for $\frac{1}{2}N-n_{\max} \le S^z < \frac{1}{2}N$
and the distortion disappears for $h<h_2$.

The origin for the different behavior
of the square-kagom\'{e} and the kagom\'{e} lattice below $h_2$
can be attributed to the circumstance,
that the square-kagom\'{e} lattice
has non-equivalent NN bonds (namely bonds belonging to triangles and
bonds belonging to squares) which carry different NN spin-spin correlations,
but the kagom\'{e} lattice has not.
This difference in the NN spin-spin correlations
leads to a special affinity of the magnetic system
to the considered lattice distortions.

Let us now briefly discuss
the influence of the lattice distortion
on the saturation field $h_1$.
Since the fully polarized ferromagnetic state is not distorted,
$h_1$ is shifted to higher values
according to
$\frac{h_1(\delta^\star)}{J}
=3+\frac{3\left(2+\sqrt{3}\right)}{32\left(6-\sqrt{3}\right)}
\frac{J}{\alpha}$
for the square-kagom\'{e}
and
$\frac{h_1(\delta^\star)}{J}
=3+\frac{1}{16}\frac{J}{\alpha}$
for the kagom\'{e} lattice,
i.e. the ``distorted magnon crystal'' remains stable
until $h_1(\delta^\star)>h_1$.
On the other hand,
starting at large magnetic fields $h > h_1$
the fully polarized ferromagnetic state remains (meta)stable
until $h_1=3J$.
Consequently,
we have a hysteresis phenomenon
in the vicinity of saturation field.

We mention that our considerations basically remain unchanged
for the anisotropic Hamiltonian (\ref{01})
with $\Delta\ne 1$.
The presence of anisotropy
leads only to quantitative changes in our results.
For the sectors of $S^z$ below the localized magnon states
we have checked that numerically for
$N=18$ (kagom\'{e})
and
$N=24$ (square-kagom\'{e}).
For the sectors of $S^z$ with localized magnon states
it becomes  obvious
from the change in magnetic energy
of a   localized magnon state due to distortions given by
$-\left(\sqrt{2}
+\frac{1}{4}\sqrt{2}\left(\sqrt{3}-1\right)\Delta\right)\delta
J$
(square-ka\-gom\'{e})
and
$-\left(1+\frac{1}{2}\Delta\right)\delta J$
(kagom\'{e}),
where these expressions for $\Delta=1$
transform to the first terms in Eqs. (\ref{07}) and (\ref{08}),
respectively.

From the experimental point of view
the discussed effect
should most spectacularly manifest itself
as a hysteresis in the magnetization and the deformation
of kagom\'{e}-lattice antiferromagnets
or the kagom\'{e}-like magnetic molecules
in the vicinity of the saturation field.

We wish to stress that the predicted 
spin-Peierls instability in high magnetic fields
may appear in a whole class 
of frustrated quantum magnets in one, two and three dimensions
hosting independent localized magnons
provided it is possible to construct
a lattice distortion preserving the symmetry of the localized-magnon cell.
Moreover, the effect is not
restricted to $s=\frac{1}{2}$ and to isotropic Heisenberg systems\cite{05,15}. 
We mention, that we have checked explicitly that our results are valid for 
e.g. the kagom\'{e}-like chain of Ref. \onlinecite{wksme}
or the dimer-plaquette chain
(see Ref. \onlinecite{schul02} and references therein).
This fact certainly increases the chance
to observe the predicted spin-Peierls instability.

There is  an increasing number of synthesized quantum frustrated kagom\'{e}
magnets\cite{ramirez00dd,19,20,bono04}.
Though these available  materials do not fit  
perfectly to an  ideal kagom\'{e} Heisenberg antiferromagnet
the physical effects based on localized magnon states may survive 
in non-ideal geometries, see  Ref. 
\onlinecite{entropy}.
Furthermore one needs comparably small exchange constants $J$
to reach experimentally the saturation field
where the structural instability occurs. A simple calculation leads to 
the relation $h_{sat}$/Tesla $\sim 2.23$ $J$/K for a
spin-$\frac{1}{2}$ kagom\'{e} Heisenberg antiferromagnet.
As  a candidate for an experimental study at saturation field may serve
the novel spin-$\frac{3}{2}$ kagom\'{e} like 
material Ba$_2$Sn$_2$ZnCr$_{7p}$Ga$_{10-7p}$O$_{22}$\cite{bono04}
with a comparably small exchange constant of about $J \sim 37 \ldots 40$K
leading to a
   saturation field accessible with today available high-end experimental
equipment. 
We may expect that further such materials will be synthesized
and with our work we are pointing out
that the efforts in this direction are worthwhile also
because of a new effect:
the spin-Peierls instability in a strong magnetic field.

To summarize,
we have reported
a spin-Peierls instability in strong magnetic fields
for several frustrated Heisenberg antiferromagnets
hosting independent localized magnons.
This spin-Peierls instability may or may not survive for smaller fields
in dependence on details of lattice structure.
In particular,
for the Heisenberg antiferromagnet on the  kagom\'{e} lattice
we have found  evidence
that
the spin-Peierls instability
breaking spontaneously the translational symmetry of the kagom\'{e}
lattice
appears only in a certain region of the magnetic field.
The field dependence of the magnetization and the deformation
in the vicinity of saturation displays a hysteresis.

We thank
A.~Honecker, J.~Schnack, S.~L.~Drechsler and D.~Ihle
for discussions and comments.
The present study was supported by the DFG
(project 436 UKR 17/17/03).

\clearpage

\noindent
{\bf Postal addresses:}\\
Prof. Dr. Johannes Richter\\
Institut f\"{u}r Theoretische Physik,
Universit\"{a}t Magdeburg,\\
P.O. Box 4120, D-39016 Magdeburg, Germany\\
Tel: +49 391 6718841 \\
Fax: +49 391 6711217 \\
E-mail: johannes.richter@physik.uni-magdeburg.de\\
Dr. Oleg Derzhko\\
Institute for Condensed Matter Physics,
National Academy of Sciences of Ukraine,\\
1 Svientsitskii Street, L'viv-11, 79011, Ukraine\\
Tel: (0322) 76 19 78\\
Fax: (0322) 76 11 58\\
E-mail: derzhko@icmp.lviv.ua\\
Dr. J\"{o}rg Schulenburg\\
Universit\"{a}tsrechenzentrum,
Universit\"{a}t Magdeburg,\\
P.O. Box 4120, D-39016 Magdeburg, Germany\\
Tel: \\
Fax: \\
E-mail: Joerg.Schulenburg@URZ.Uni-Magdeburg.DE


\begin{thebibliography}{99}

\bibitem{01}
C.~Lhuillier and G.~Misguich,
in
Lecture Notes in Physics {\bf 595}
(Springer, Berlin, 2002),
pp.161-190

\bibitem{02}
G.~Misguich and C.~Lhuillier,
arXiv:cond-mat/0310405

\bibitem{waldtmann98}
Ch.~Waldtmann et al., 
Eur. Phys. J. B {\bf 2}, 501 (1998)

\bibitem{CuGeO}
M.~Hase et al.,
Phys. Rev. Lett. {\bf 70,} 3651 (1993)

\bibitem{03}
R.~Coldea et al.,
Phys. Rev. B {\bf 68,} 134424 (2003)



\bibitem{04}
A.~Honecker et al.,
J. Phys.: Condens. Matter {\bf 16}, S749 (2004)

\bibitem{nojiri88}
H.~Nojiri et al.,
J. Phys. Colloq. France {\bf 49,} C8-1459 (1988)

\bibitem{kageyama99}
H.~Kageyama et al.,
Phys. Rev. Lett. {\bf 82,} 3168 (1999)

\bibitem{majumdar_ghosh}
C.~K.~Majumdar and D.~K.~Ghosh,
J. Math. Phys. {\bf 10,} 1388 (1969)

\bibitem{ivanov_richter}
N.~B.~Ivanov and J.~Richter,
Phys. Lett. A {\bf 232,} 308 (1997)

\bibitem{05}
J.~Schulenburg et al., 
Phys. Rev. Lett. {\bf 88,} 167207 (2002);
J.~Richter et al.,
J. Phys.: Condens. Matter {\bf 16}, S779 (2004)

\bibitem{07}
F.~Becca, F.~Mila, and D.~Poilblanc,
Phys. Rev. Lett. {\bf 91,} 067202 (2003)

\bibitem{08}
F.~Becca and F.~Mila,
Phys. Rev. Lett. {\bf 89,} 037204 (2002)

\bibitem{09}
P.~Carretta et al.,
Phys. Rev. B {\bf 66}, 094420 (2002)

\bibitem{10}
S.~-~H.~Lee et al., 
Phys. Rev. Lett. {\bf 84,} 3718 (2000)

\bibitem{11}
A.~Keren and J.~S.~Gardner,
Phys. Rev. Lett. {\bf 87,} 177201 (2001)

\bibitem{12}
Y.~Yamashita and K.~Ueda,
Phys. Rev. Lett. {\bf 85,} 4960 (2000)

\bibitem{13}
O.~Tchernyshyov et al., 
Phys. Rev. Lett. {\bf 88,} 067203 (2002)

\bibitem{gross}
M.~C.~Gross,
Phys. Rev. B {\bf 20,} 4606 (1979)

\bibitem{14}
R.~Siddharthan and A.~Georges,
Phys. Rev. B {\bf 65,} 014417 (2001)

\bibitem{tomczak}
P.~Tomczak and J.~Richter,
J. Phys. A {\bf 36}, 5399 (2003)

\bibitem{15}
J.~Schnack et al., 
Eur. Phys. J. B {\bf 24,} 475 (2001);
H.~-~J.~Schmidt,
J. Phys. A {\bf 35,} 6545 (2002)

\bibitem{17}
T.~Momoi and K.~Totsuka,
Phys. Rev. B {\bf 61,} 3231 (2000)

\bibitem{18}
M.~Oshikawa,
Phys. Rev. Lett. {\bf 84,} 1535 (2000)

\bibitem{wksme}
Ch.~Waldtmann et al.,
Phys. Rev. B {\bf 62,} 9472 (2000)

\bibitem{schul02}
J.~Schulenburg and  J.~Richter,
Phys. Rev. B {\bf 66,} 134419 (2002)




\bibitem{ramirez00dd}
A.~P.~Ramirez et al.,
Phys. Rev. Lett. \textbf{84}, 2957 (2000)



\bibitem{19}
Z.~Hiroi et al., 
J. Phys. Soc. Jpn. {\bf 70,} 3377 (2001)

\bibitem{20}
A.~Fukaya et al., 
Phys. Rev. Lett. {\bf 91,} 207603 (2003)



\bibitem{bono04} D.~Bono et al., 
 {Phys. Rev. Lett.} \textbf{92}, 217202 (2004)


\bibitem{entropy} 
O.~Derzhko and  J.~Richter, arXiv:cond-mat/0404204

\end{thebibliography}
\end{document}